\documentclass[%
superscriptaddress,
reprint,
amsmath,amssymb,
pra,
noeprint,
]{revtex4-2}

\usepackage[pdftex]{graphicx}
\graphicspath{{./fig/}{C:/Users/scott/OneDrive/document/kosakaLab/paper/CoherentControl_NV0/tex/fig/}{./}}

\usepackage{braket}
\usepackage{here}

\usepackage{comment}
\usepackage{dcolumn}
\usepackage{bm}
\usepackage{hyperref}
\usepackage{amsmath}
\usepackage{color}

\begin{document}
	
	\title{Coherent Electric Field Control of Orbital State of a Neutral Nitrogen-Vacancy Center}
	
	
	\author{Hodaka Kurokawa}
	\email[E-mail: ]{kurokawa-hodaka-hm@ynu.ac.jp}
	\address{Quantum Information Research Center, Institute of Advanced Sciences, Yokohama National University, 79-5 Tokiwadai, Hodogaya, Yokohama 240-8501, Japan}
	\author{Keidai Wakamatsu}
	\address{Department of Physics, Graduate School of Engineering Science, Yokohama National University,  79-5 Tokiwadai, Hodogaya, Yokohama 240-8501, Japan }
	\author{Shintaro Nakazato}
	\address{Department of Physics, Graduate School of Engineering Science, Yokohama National University,  79-5 Tokiwadai, Hodogaya, Yokohama 240-8501, Japan }
	\author{Toshiharu Makino}
	\address{Quantum Information Research Center, Institute of Advanced Sciences, Yokohama National University, 79-5 Tokiwadai, Hodogaya, Yokohama 240-8501, Japan}
	\address{Advanced Power Electronics Research Center, National Institute of Advanced Industrial Science and Technology, 1-1-1 Umezono, Tsukuba, Ibaraki, 305-8568, Japan }
	\author{Hiromitsu Kato}
	\address{Quantum Information Research Center, Institute of Advanced Sciences, Yokohama National University, 79-5 Tokiwadai, Hodogaya, Yokohama 240-8501, Japan}
	\address{Advanced Power Electronics Research Center, National Institute of Advanced Industrial Science and Technology, 1-1-1 Umezono, Tsukuba, Ibaraki, 305-8568, Japan }
	\author{Yuhei Sekiguchi}
	\address{Quantum Information Research Center, Institute of Advanced Sciences, Yokohama National University, 79-5 Tokiwadai, Hodogaya, Yokohama 240-8501, Japan}
	\author{Hideo Kosaka}
	\email[E-mail: ]{kosaka-hideo-yp@ynu.ac.jp}
	\address{Quantum Information Research Center, Institute of Advanced Sciences, Yokohama National University, 79-5 Tokiwadai, Hodogaya, Yokohama 240-8501, Japan}
	\address{Department of Physics, Graduate School of Engineering Science, Yokohama National University,  79-5 Tokiwadai, Hodogaya, Yokohama 240-8501, Japan }
		
\begin{abstract}
The coherent control of the orbital state is crucial for realizing the extremely-low power manipulation of the color centers in diamonds. Herein, a neutrally-charged nitrogen-vacancy center, NV$^0$, is proposed as an ideal system for orbital control using electric fields. The electric susceptibility in the ground state of NV$^0$ is estimated, and found to be comparable to that in the excited state of NV$^-$. Also, the coherent control of the orbital states of NV$^0$ is demonstrated. The required power for orbital control is three orders of magnitude smaller than that for spin control, highlighting the potential for interfacing a superconducting qubit operated in a dilution refrigerator.

\end{abstract}
\maketitle
	
Diamond color centers are attracting increased attention because of their potential applications in quantum communication \cite{Bernien2013,Pompili2021,Bhaskar2020}, quantum computation \cite{Nemoto2014,Abobeih2019}, and quantum sensing \cite{Dolde2011,Degen2017}. The spin degree of freedom is primarily utilized as a quantum bit owing to its long coherence time of over one second \cite{Balasubramanian2009,Bar-Gill2013,Abobeih2018}, and excellent controllability \cite{Abobeih2022, Nguyen2019}. However, the control of the orbital degree of freedom is also crucial for various applications such as the frequency tuning of the zero-phonon line photons and extremely low-power control of the electron states. The ability to tune the zero-phonon line frequency through electric fields or strain is essential for generating entanglement between remote color centers \cite{Tamarat2008,Maze2011,Bernien2013}. Moreover, the coupling of electric fields or strain with the orbital degree of freedom is stronger compared to the magnetic field's coupling with the spin \cite{Meesala2018,Chen2018a,Block2021a}, allowing the highly-efficient control of the electron state. Owing to the strong spin-orbit coupling, efficient spin-state control using the strain has been achieved in color centers \cite{Maity2020}, which is particularly advantageous for operations in a dilution refrigerator. Nevertheless, directly achieving coherent control over the orbital state remains challenging for representative color centers owing to the short lifetime of the optically-excited state of NV$^-$ ($\sim 10$ ns) \cite{Batalov2008a} and large ground-state splitting of group-IV color centers \cite{Wolfowicz2021}. 

Therefore, a neutrally-charged nitrogen-vacancy center, NV$^0$ \cite{Manson2005,Gali2009,Hauf2011,Grotz2012,Siyushev2013, Doi2014,Barson2019,Baier2020,Zhang2021,Aslam2013,Poem2015a}, is proposed herein as an ideal system for orbital state control using electric fields. The ground-state spin-orbit splitting of NV$^0$ is approximately 10 GHz \cite{Barson2019,Baier2020}, allowing direct microwave access. Additionally, the ground state of NV$^0$ exhibits an orbital relaxation time of several hundreds of nanoseconds \cite{Baier2020}, which is more than one order of magnitude longer than that of the excited state of NV$^-$. The energy level structure of the ground state in NV$^0$ is also similar to that of group-IV color centers. Therefore, NV$^0$ can serve as an ideal system for understanding the properties of the orbital state and conducting proof-of-principle experiments relevant to group-IV color centers. In this study, the electric susceptibility of NV$^0$ is investigated, and the coherent control of its orbital state is demonstrated. The highly-efficient control of the electron state creates possibilities for future applications, such as in a quantum interface communicating with a superconducting quantum bit in a dilution refrigerator \cite{Neuman2021,Kurokawa2022}.

\section*{Results}
\subsection*{Characterization of NV$^0$}

\begin{figure*}
	\begin{center}
		\includegraphics[width=180 mm]{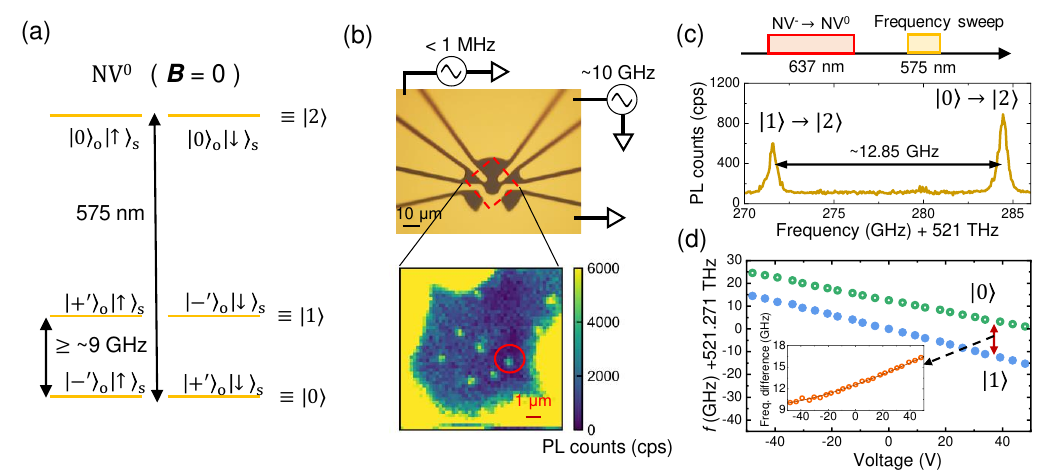}
	\end{center}
	\caption{Energy level structure of NV$^0$, device image, photoluminescence excitation (PLE) spectrum, and DC voltage dependence of the PLE spectrum. (a) Schematic of the energy level of NV$^0$ with some static strain under a zero magnetic field. The ground state exhibits energy splitting in the order of 10 GHz owing to the combined effects of spin-orbit interaction and strain. Moreover, the optical transition occurs at 575 nm. (b) Optical microscope image of the electrical circuit on the diamond used in the experiments. The upper electrodes are used to apply low ($<$ 1 MHz) or high ($\sim$10 GHz) electric fields. Moreover, the color map shows the photoluminescence (PL) counts around the center of the circuit. The red-circled NV center is used for experiments. (c) PLE spectrum of NV$^0$ and its measurement sequence. The 637-nm laser resonant at the zero-phonon line of NV$^-$ is used to initialize the charge state to NV$^0$, and the 575-nm laser is used to observe the transition. (d) PLE spectrum shifts as a function of the DC voltage. The green and blue circles correspond to the lower ($\ket{0}$) and upper ($\ket{1}$) branches, respectively, and the inset shows the difference in the PLE frequencies of the upper ($\ket{0}$) and lower $\ket{1}$ branches. Additionally, the horizontal axis is the same as that of the main graph. 
	}
	\label{fig:schematic}
\end{figure*}

Under a zero magnetic field, the ground-state Hamiltonian, $H$, of NV$^0$ in the $\ket{\pm}_{\text{o}}$ basis can be modeled as \cite{Barson2019,Baier2020}:
\begin{equation}\label{eq:basic}
	H^{(\pm)}/h=2\lambda \hat{L}_z \hat{S}_z + \epsilon_{\perp}(\hat{L}_+ +\hat{L}_-),
\end{equation}
where $h$ is the Plank's constant, $\lambda$ is the spin-orbit interaction parameter, $\hat{L}_z=\sigma_z$, and $\hat{L}_{\pm}=\ket{\pm}_{\text{o}}\bra{\mp}_{\text{o}}$ are the orbital operators in the $\ket{\pm}_{\text{o}}=\mp1/\sqrt{2}(\ket{e_x}_{\text{o}}\pm i\ket{e_y}_{\text{o}})$ basis, $\ket{e_x}_{\text{o}}$ and $\ket{e_y}_{\text{o}}$ are the strain eigenstates, $S_z=(1/2)\sigma_z$ is the spin operator for 1/2 spin, and $\epsilon_{\perp}$ is the perpendicular strain parameter. Moreover, the subscript, o, denotes the orbital degree of freedom. Figure \ref{fig:schematic} (a) shows the energy level of NV$^0$. The ground-state splitting originates from the spin-orbit coupling, $\lambda$, and the strain, $\epsilon_{\perp}$, as  $2\sqrt{\lambda^2+\epsilon_{\perp}^2}$. Under strain, the eigenstates change from $\ket{\pm}_{\text{o}}$ to $\ket{+'}_{\text{o}}=\alpha\ket{+}_{\text{o}}+\beta\ket{-}_{\text{o}}$ and $\ket{-'}_{\text{o}}=-\beta\ket{+}_{\text{o}}+\alpha\ket{-}_{\text{o}}$ with $\alpha^2+\beta^2=1$ (See Supplementaly eqs. (9)(10)). Figure \ref{fig:schematic} (b) shows the electrical circuit directly formed on a diamond. Low- ($<$ 1 MHz) and high-frequency ($\sim$10 GHz) electric fields are applied to the nitrogen vacancies using the upper electrodes. The color map of the photoluminescence (PL) counts around the center of the circuit is also shown. Figure \ref{fig:schematic} (c) shows the photoluminescence excitation (PLE) spectrum with the experimental sequence. A 637-nm red laser resonant to the zero-phonon line (ZPL) of NV$^-$ is used to convert NV$^-$ to NV$^0$. The frequencies of a 575-nm yellow laser are then swept to search for the ZPL of NV$^0$. Consequently, two transition lines are observed, which indicates that the ground-state splitting is 12.85 GHz. 

Since all experiments are performed under an ambient magnetic field, the transitions from $\ket{+'}_\mathrm{o}\ket{\uparrow}_\mathrm{s}$ and $\ket{-'}_\mathrm{o}\ket{\downarrow}_\mathrm{s}$ are indistinguishable, where the subscript, s, denotes the spin degree of freedom. Additionally, $\ket{-'}_\mathrm{o}\ket{\uparrow}_\mathrm{s}$ and $\ket{+'}_\mathrm{o}\ket{\downarrow}_\mathrm{s}$ are indistinguishable. These states can be rewritten as $\ket{-'}_\mathrm{o}\ket{\uparrow}_\mathrm{s}=\ket{+'}_\mathrm{o}\ket{\downarrow}_\mathrm{s}\equiv\ket{0}$, $\ket{+'}_\mathrm{o}\ket{\uparrow}_\mathrm{s}=\ket{-'}_\mathrm{o}\ket{\downarrow}_\mathrm{s}\equiv\ket{1}$, and $\ket{0}_\mathrm{o}\ket{\uparrow}_\mathrm{s}=\ket{0}_\mathrm{o}\ket{\downarrow}_\mathrm{s}\equiv\ket{2}$ by ignoring the spin degree of freedom. In other words, the four-level system in the ground state is treated as effective two-level systems with different spin states. 
This assumption holds for spin-independent experiments or experiments with time scales shorter than spin relaxation time. Basically, our experiments are assumed to be spin-independent because we can drive and readout both spin states optically and electrically. Furthermore, although reported spin relaxation times range from 570 $\mu$s to 1.5 s depending on situations \cite{Baier2020, Bradley2022}, they are still longer than the timescale of our experiments ($\sim$ a few $\mu$s).

Subsequently, DC electric fields are applied to estimate the parameters in eq. (\ref{eq:basic}) and the electric susceptibility.
The effects of both the strain and DC electric fields in the $\ket{\pm}_{\text{o}}\ket{\uparrow}_{\text{s}}$ basis can be expressed as:
\begin{equation}
	\begin{split}
			H^{(\pm)}/h = \lambda \hat{L}_z + d_{\parallel}E_z\hat{I}+ (\epsilon_{\perp}+d_{\perp}E_{\perp})(\hat{L}_+ + \hat{L}_-) \\ 
			+d_{\perp}E_{\perp}'(-i\hat{L}_+ + i\hat{L}_-),		
	\end{split}
\end{equation}
	where $d_{\parallel}$ ($d_{\perp}$) is the electric susceptibility parallel (perpendicular) to the NV axis, $E_z$ is the electric fields parallel to the NV axis, $E_{\perp}$ is the electric fields whose axis is parallel to the direction of $\epsilon_{\perp}$, and $E_{\perp}'$ is the electric fields perpendicular to $E_{\perp}$ and $E_z$. Here, the strain term which only contributes to the global shift of the energy is ignored (See Supplementary section I). The energy eigenvalues are $d_{\parallel}E_z\pm\sqrt{\lambda^2+(\epsilon_{\perp}+E_{\perp})^2+(d_{\perp}E_{\perp}')^2}$. Figure \ref{fig:schematic} (d) shows the PLE frequency shifts at both the upper ($\ket{1}$) and lower ($\ket{0}$) branches as a function of the DC electric fields. With increasing DC electric fields, the PLE frequencies of the two branches decrease almost linearly. Additionally, with the application of $\pm$50 V, the PLE frequencies can be shifted up to $\sim$30 GHz. The linear frequency shifts in both branches are caused by $E_z$. Using the electric field distribution obtained from the finite element simulation, $d_{\parallel}$ is estimated to be 1.08 MHz/(V cm$^{-1}$) (See Supplementary section II), which is similar to the reported value in the excited state of NV$^-$ of 0.7 MHz/(V cm$^{-1}$) \cite{Block2021a}. The frequency shifts owing to $E_{\perp}$ and $E_{\perp}'$ can be clearly seen by subtracting the upper PLE frequencies from the lower PLE frequencies (Inset of Fig. \ref{fig:schematic} (d)). The frequency differences are then fitted using the equation $2\sqrt{\lambda^2+(\epsilon_{\perp}+d_{\perp}E_{\perp})^2+(d_{\perp}E_{\perp}')^2}$ and a fixed $\lambda=4.80$ GHz \cite{Baier2020}. Consequently, the estimated parameters are $\epsilon_{\perp}$ = 4.06 GHz and $d_{\perp}$ = 348 (12) kHz/(V cm$^{-1}$). It should be noted that $d_{\perp}$ is several times smaller than the value reported in Ref. \cite{Block2021a} for the excited state of NV$^{-}$, 1.4 MHz/(V cm$^{-1}$).

\begin{figure}
	\begin{center}
		\includegraphics[width=80 mm]{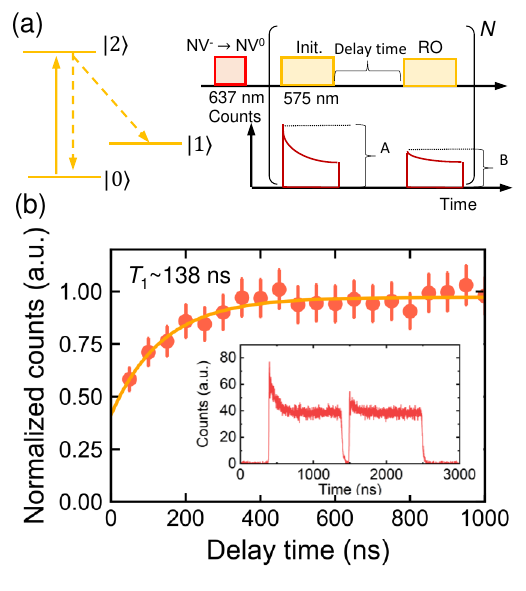}
	\end{center}
	\caption{Measurement of the orbital relaxation time of NV$^0$. (a) Schematic of the experimental sequence for measuring the orbital relaxation time of NV$^0$. After charge initialization (Init.) using the 637-nm laser, the first 575-nm laser pulse initializes $\ket{0}$ to $\ket{1}$, and the population in $\ket{0}$ is read out (RO) by the second 575-nm pulse after some delay time. The peak counts, $A$ and $B$, at each pulse are used to estimate the population in $\ket{0}$. The repetition time for each measurement, $N$, is 150, and is determined by measuring the decay of the PL counts as a function of the duration of the 575-nm laser irradiation. (b) Normalized counts, $B/A$, as a function of the delay time. The red dots are the experimental data and the orange curve is the results of curve fitting using the exponential function. The inset shows the time-resolved PL counts of two pulses with a 100 ns gap in between. Additionally, the PL counts are smoothed using the Savitzky-Golay filter with 17 points before the calculation of the peak height, and the background is subtracted using the data when the laser is absent. Error bars correspond to standard deviation error after the smoothing.
	}
	\label{fig:T1}
\end{figure}

\subsection*{Orbital relaxation of NV$^0$}
The orbital relaxation time, $T_1$, is important because it determines the limit of the operation time. Figure \ref{fig:T1} (a) shows the experimental sequence for the measurement of $T_1$. A 1-$\mu$s long optical pulse resonant is applied to the transition between $\ket{0}$ and $\ket{2}$ to initialize NV$^0$ into $\ket{1}$ from the thermal mixture of $\ket{0}$ and $\ket{1}$ (Inset of Fig. \ref{fig:T1} (b)). The first peak height is proportional to the initial population in $\ket{0}$, and the peak height in the second pulse is proportional to the population in $\ket{0}$ relaxed from $\ket{1}$. The ratio of the pulse heights in the first and second pulses is a measure of the decay from $\ket{1}$, and was used to calculate $T_1$. 
Figure \ref{fig:T1} (b) shows the normalized pulse height as a function of the delay of the second pulse. $T_1$ is estimated to be $\sim138$ (19) ns at 5.5 K based on the curve fit with $1-a\mathrm{exp}(-t/T_1)+b$, where $a$ and $b$ are constants. In Ref. \cite{Baier2020}, $T_1$ is a few times larger than ours. Since $T_1$ is limited by thermal phonons at 5.5 K, a further increase in $T_1$ is expected by lowering the temperature to several tens of millikelvin.

\begin{figure}
	\begin{center}
		\includegraphics[width=80 mm]{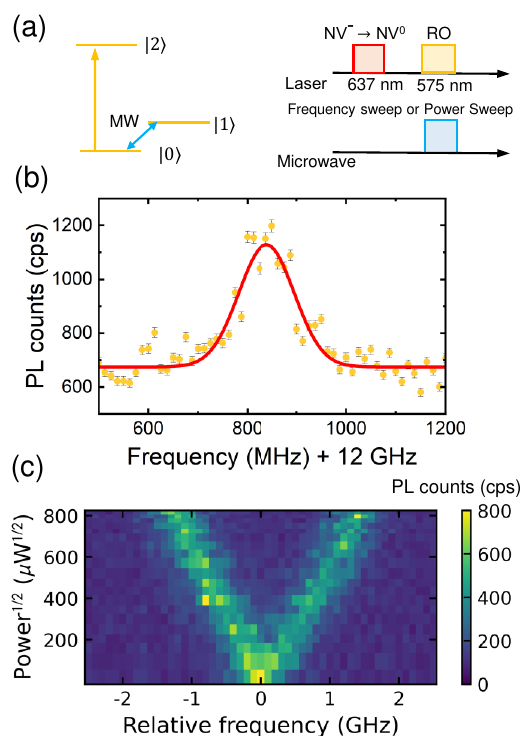}
	\end{center}
	\caption{Optically-detected electrical resonance (ODER) and Rabi splitting measurements. (a) Schematic of the experimental sequence used in measuring the ODER and Rabi splitting. After charge conversion using the 637-nm laser, the 575-nm laser is simultaneously applied with microwave application at the transition between $\ket{0}$ and $\ket{2}$. The frequencies of the 575-nm laser are then swept for ODER measurement, and the power of the microwave is swept for the measurement of Rabi splitting. (b) ODER spectrum of NV$^0$. The orange dots represent the data and the red curve is the Gaussian fit. (c) The Autler-Townes splitting as a function of the square root of the power of the applied microwave electric fields. 
	}
	\label{fig:ODER_ACstark}
\end{figure}

\subsection*{Optically-detected electrical resonance and Rabi splitting}

 To investigate the electrical resonance frequency in the ground state of NV$^0$, the optically-detected electrical resonance (ODER) is employed. Since the driving electric field is resonant to the eigenfrequency determined by $\lambda$ and $\epsilon_{\perp}$, the AC driving electric fields is treated as a perturbation to $\ket{\pm'}_{\text{o}}$ which is the eigenstates of eq. (\ref{eq:basic}).
 In the rotating frame of the driving AC electric fields, $H_{\mathrm{d}}/h=f_{\mathrm{d}}\hat{L}_z/2$, under the rotating wave approximation, the Hamiltonian of NV$^0$ in the $\ket{\pm'}_{\textrm{o}}$ basis can be written as:
 \begin{equation}\label{Hamiltonian-electric}
 	H^{(\pm')}/h=\frac{\Delta}{2} \hat{L}_z + \frac{d_{\perp}E_{\perp}''}{2}(\hat{L}_+ + \hat{L}_-),
 \end{equation}
 where $f_\text{d}$ is the frequency of the driving electric fields, $\Delta=2\sqrt{\lambda^2+\epsilon_{\perp}^2}-f_{\mathrm{d}}$ is the detuning, $E_{\perp}''=\sqrt{(\alpha^2-\beta^2)E_x^2+E_y^2}$ represents the electric fields perpendicular to the NV axis, and $E_x$ and $E_y$ are the in-plane electric fields (See Supplementary section I). Here, $\hat{L}_{\pm}$ are the rising (lowering) operators in the $\ket{\pm'}_{\text{o}}$ basis. When $\Delta\sim0$, the second term that remains in eq. (\ref{Hamiltonian-electric}) contributes to Rabi splitting and Rabi oscillation, which will be discussed in below.

 Figure \ref{fig:ODER_ACstark} (a) shows the experimental sequence. The microwave frequencies are swept around the frequency resonant to the transition between $\ket{0}$ and $\ket{1}$, which is roughly estimated from the PLE measurement. Simultaneously, the readout laser is applied to measure the population in $\ket{0}$. A sufficiently long readout pulse increases the population in $\ket{1}$ through optical pumping, as shown in the inset of Fig. \ref{fig:T1} (b). When the microwave frequency and ground-state splitting are in resonance, the population in $\ket{1}$ is transferred to $\ket{0}$, resulting in an increase in the PL counts. Figure \ref{fig:ODER_ACstark} (b) shows the results of the ODER measurement. The maximum of the spectrum is 12.84 GHz, which is in good agreement with the ground-state splitting obtained from the PLE measurement. Additionally, the full width at half maximum (FWHM) is 133 (10) MHz due to the spectral diffusion in the ground and optically-excited states.

 Using the resonance frequency (12.84 GHz) obtained from the ODER measurement, the microwave power dependence of the Autler-Townes splitting is investigated by increasing the microwave power. The Autler-Townes splitting originates from the formation of the color center-field dressed states because of the strong driving field (See Supplementary section I). The splitting between the two peaks corresponds to $d_{\perp}E_{\perp}''$, the Rabi frequency. Figure \ref{fig:ODER_ACstark} (c) shows the color map of the PL counts as functions of the relative frequency and square root of the microwave power. The maximum Autler-Townes splitting is 2.6 GHz at 824 $\mu$W$^{1/2}$. The power dependence of Autler-Townes splitting directly corresponds to the power dependence of the Rabi frequency, which is estimated to be 3.37 MHz/$\mu$W$^{1/2}$ via fitting.
 Additionally, the electric susceptibility is estimated to be $d_{\perp}^{\mathrm{AC}}=$ 1.0 MHz/(V cm$^{-1}$) (See Supplementary section II). Here, $d_{\perp}^{\mathrm{AC}}$ is larger than that obtained from the measurement using DC electric fields (348 kHz/(V cm$^{-1}$)). The difference can be attributed to electric field screening during DC measurement, which decreases the effective electric fields at the NV center \cite{Block2021a}. The value of $d_{\perp}^{\mathrm{AC}}$ is in relatively good agreement with that of the optically-excited state of NV$^-$, which is 1.4 MHz/(V cm$^{-1}$) \cite{Block2021a}.

 \begin{figure}
 	\begin{center}
 		\includegraphics[width=80 mm]{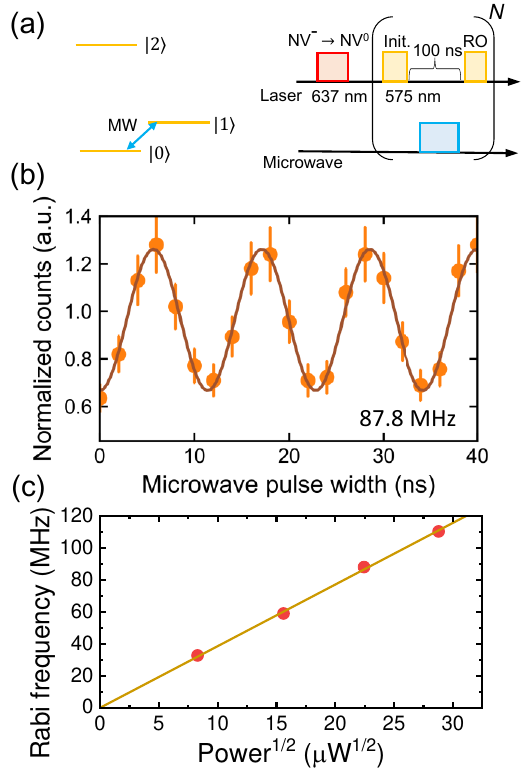}
 	\end{center}
 	\caption{Rabi oscillation and microwave power dependence of the Rabi frequency. (a) Schematic of the experimental sequence for Rabi oscillation. The microwave pulse is applied between the first (initialization) and second (readout) laser pulses, and the microwave frequency is set to the resonance frequency obtained from ODER measurement. The repetition times, $N$, are 100. (b) Normalized counts as a function of the microwave pulse width. The orange dots represent the data and the brown curve is the cosinusoidal fit. Error bars correspond to standard deviation error. The Rabi frequency is 87.8 MHz. (c) Rabi frequency as a function of the square root of the input power. The slope is 3.86 MHz/$\mu$W$^{1/2}$.}
 	\label{fig:Rabi}
 \end{figure}

\subsection*{Rabi oscillation and Ramsey interference}
Using the resonance frequency obtained from ODER measurement, the Rabi oscillations between $\ket{0}$ and $\ket{1}$ are observed. Figure \ref{fig:Rabi} (a) shows the experimental sequence for measurement, where the microwave pulse is applied between the two laser pulses. Figure \ref{fig:Rabi} (b) shows the Rabi oscillation at an input microwave power of 504 $\mu$W, where the Rabi frequency is 87.8 MHz. Compared to the power required to drive a spin of NV$^-$ at a similar distance from the electrode (several $\mu$m), the required power is three orders of magnitude smaller. Figure \ref{fig:Rabi} (c) shows the Rabi frequency as a function of the square root of the microwave power. Here, the Rabi frequency linearly increases with the square root of the microwave power, and the slope is 3.86 (3)  MHz/$\mu$W$^{1/2}$, which is in good agreement with the value obtained from the Rabi splitting measurement of 3.37 MHz/$\mu$W$^{1/2}$.

 \begin{figure}
	\begin{center}
		\includegraphics[width=80 mm]{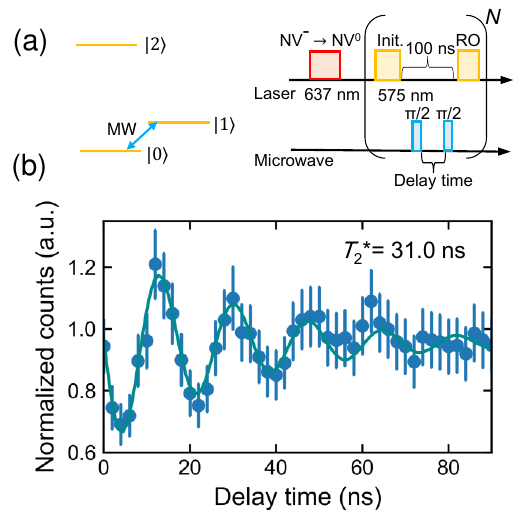}
	\end{center}
	\caption{Ramsey interference. (a) Schematic of the experimental sequence for Ramsey interference, where the two microwave pulses are applied between the first (initialization) and second (readout) laser pulses. The microwave width for the $\pi/2$ pulse is determined to be 4.6 ns based on the Rabi oscillation measurement. The repetition times, $N$, are 100. (b) Normalized counts as a function of the delay time of the microwave pulse (free precession time). The blue dots represent the data and the blue curve represents the fitting using the function, $A\mathrm{sin}(\Delta t+\phi)\mathrm{exp}(-t/T_2^*)+B$, where $A$, $B$, and $\phi$ are constants, $\Delta$ is the detuning from the transition between $\ket{0}$ and $\ket{1}$, and $T_2^*$ is the orbital coherence time. Error bars correspond to standard deviation error. $T_2^*$ is 31.0 ns.}
	\label{fig:Ramsey}
\end{figure}
 
 The coherence time, $T_2^*$, of the orbital state is also investigated using the Ramsey interference. Figure \ref{fig:Ramsey} (a) shows the experimental sequence. The two microwave $\pi/2$ pulses are applied between the initializing and readout laser pulses. Figure \ref{fig:Ramsey} (b) shows the Ramsey interference with some detuning as a function of the free precession time. Based on curve fitting, the detuning is 58 MHz and $T_2^*$ is 31.0 (3.6) ns. The origins of the decoherence are attributed  spectral diffusion \cite{Robledo2010} in the ground state  and thermal phonons that are resonant with the orbital transition proposed for SiV \cite{Jahnke2015}. The value of $T_2^*$ is similar to the spin coherence time of SiV at approximately 5 K (30-50 ns) \cite{Rogers2014a,Pingault2014,Maity2020}. Thus, a further increase in $T_2^*$ is expected by decreasing the temperature and the application of the dynamical decoupling as SiV.

\section*{Discussion}
The electric susceptibility of NV$^0$ has been measured in the ground state, and found to be comparable to that in the excited state of NV$^-$. Additionally, the coherent control of the orbital state of NV$^0$ has been achieved using electric fields. The required power for Rabi oscillation is hundreds of microwatts, which is three orders of magnitude smaller than that required for spin control using magnetic fields (See Supplementary section III). 

The highly-efficient control of the orbital state is particularly advantageous for the operations in a dilution refrigerator, opening up the possibility for interfacing a superconducting qubit and a color center via electric fields. If a high impedance superconducting microwave resonator with a zero-point voltage fluctuation of tens of microvolts \cite{Samkharadze2016,Niepce2019} is integrated, the single-photon coupling is expected to be tens of kilohertz, allowing further low-power microwave control using several hundred microwave photons. 
If we can further increase the coupling to $\sim$1 MHz by bringing the electrode and the color center closer together, we can reach the strong coupling regime of the color center and the microwave resonator, where we can observe single-photon interaction.

Also, we should compare the electric field control of the  orbital state with another driving scheme, mechanical driving \cite{Maity2020}. The electric field driving requires less complex devices and offers wide frequency  range while typical mechanical oscillators employed in experiments are mechanical resonators with narrow frequency range around resonant frequency \cite{Chen2018a,Maity2020}. However, though device fabrication processes and frequency tuning result in some more complexity in experiments, controlling the orbital state using phonons is still promising. Since the mode volume of the phononic resonator can be designed to be several orders of magnitude smaller than that of typical microwave resonators \cite{Neuman2021,Kim2023}, the single-phonon coupling can exceed 1 MHz without the use of extremely close proximity electrodes as mentioned above, allowing to reach the strong coupling regime with maintaining both microwave and optical access.


In addition, a comparison with group IV color centers needs to be discussed. Although the ground state can be modeled with the same Hamiltonian for NV$^0$ and group IV color centers, considering the spin-orbit coupling, the spin (orbital) Zeeman effect, the strain, and the Jahn-Teller effect which is indistinguishable from strain, the coefficients in each term are different. The largest difference in parameters would be the spin-orbit coupling, which ranges from $\sim$ 5 GHz in NV$^0$ to $\sim$ 400 GHz (corresponding to the splitting of $\sim$ 800 GHz) in PbV. The ratio of spin-orbit coupling to strain determines the degree of hybridization of the ground state wavefunction, which is essential for magnetic field control of the ground state spin. For NV$^0$ with 5 GHz spin-orbit coupling, several GHz to tens of GHz of strain parameters are sufficient to fully tune its wavefunction from the spin-orbit coupling dominant regime to the strain dominant regime. In contrast, tens to hundreds of GHz of strain are required to fully control the wavefunction of group IV color centers. Furthermore, we can control the Hamiltonian of NV$^0$ using dc electric fields, effectively suppressing or enhancing the effect of static strain. Thus, NV$^0$ can be a good platform to study the effect of strain on the Hamiltonian in a wide parameter range with relatively small strain parameters and dc electric fields. It should be noted that the excited state Hamiltonian of NV$^0$ and group IV color centers are different. While NV$^0$ has two optically excited states with two spin degrees of freedom and one orbital degree of freedom, group IV color centers have optically excited states with two spin degrees of freedom and two orbital degrees of freedom. This difference may affect the readout scheme when considering the single-photon readout demonstrated in Ref. \cite{Nguyen2019}.


Although the measured $T_1$ is 138 ns at 5.5 K, it is expected to extend to microseconds in the dilution refrigerator environment due to a decrease in the thermal phonons. If it reaches several microseconds, it can be used as an interface between the superconducting quantum bit \cite{Kurokawa2022}, which is proposed for the silicon-vacancy center \cite{Neuman2021}. Despite of the several challenges posed by spectral diffusion, the integration of the microwave resonator, and operations in the dilution refrigerator, this study demonstrates a promising path for hybrid quantum systems.

\section*{Methods}
\subsection*{Sample fabrication}
 [100]-cut electronic-grade single-crystal diamond samples are synthesized via chemical vapor deposition (CVD) (element six). Before the fabrication of the electrode, the diamond substrate is kept in a mixture of H$_2$SO$_4$ and HNO$_3$ at 200${}^\circ$C for 60 min to remove any surface contamination and terminate the surface with oxygen. Subsequently, Au (500 nm)/Ti (10nm) electrodes are formed on the substrate through photolithography processes.

\subsection*{Low-temperature confocal microscopy}
All experiments are performed in a closed-cycle optical cryostat (Cryostation s50, Montana Instruments) at 5.5 K under an ambient magnetic field. 
A copper sample holder is mounted on the XYZ piezoelectric nanopositioners (ANPx101$\times2$, ANPz101) using a thermal link, and the diamond sample is fixed onto a homemade printed circuit board (PCB) using aluminum tapes. Additionally, the electrodes on the diamond are connected to the PCB using gold wire bonds, and the PCB is attached to the sample holder using screws.

Optical excitation and collection are then performed using a home-built confocal microscope. An objective lens (LMPLFLN100X, Olympus), whose numerical aperture is 0.8, is scanned over the sample using a XYZ piezoelectric nanopositioner (P-517.3CD, Physik Instrumente). 

A 515-nm green laser (Cobolt, Hubner Photonics) is used for off-resonant excitation to search for the nitrogen-vacancy centers, and a 575-nm yellow laser (DL-SHG pro, Toptica Photonics) is used to resonantly excite NV$^0$ for initialization and readout. Typically, a 1-$\mu$s pulse is used and the power is 1 $\mu$W. A 637-nm red laser (DL pro, Toptica Photonics) is also used to resonantly excite NV$^-$ to convert it from the negatively-charged state to the neutrally-charged state. The pulse width for charge initialization is 100 $\mu$s and the power is 200 $\mu$W. 
The charge initialization fidelity is estimated to be $>97.0\%$ from the residual fluorescence after the irradiation of long charge initialization pulse.
Each laser is directed along the main optical path using dichroic mirrors (FF552-Di02-25x36: 515 nm, FF605-Di02-25x36: 575 nm, FF649-Di01-25x36: 637 nm, Semrock) after being purified through bandpass filters (FF02-520/28-25: 515 nm, FF03-575/25-25: 575 nm, FF01-637/7-25: 637 nm, Semrock). 
 
The power of the 515-nm laser is controlled by setting the internal output power, and those of the 575-nm and 637-nm lasers are controlled using variable fiber-optic attenuators (V450PA: 575 nm, V600PA: 637 nm, Thorlabs). Additionally, the pulse duration of the 515-nm laser is directly set based on the signal sent from a field programmable gate array (FPGA).
The pulse durations of the 575-nm and 637-nm laser are gated using acousto-optic modulators (SFO4903-S-M200 0.4C2C-3-F2P-01: 575 nm, SFO3916-S-M200-0.4C2E-3-F2P-02: 637 nm, Gooch and Housego) controlled by the FPGA (PXIe-7820R, National Instruments). 

The phonon sideband of NV$^0$ and NV$^-$ is collected and measured using a single-photon counter (SPCM-AQRH-14-FC, Excelitas). For the time-resolved photoluminescence measurements, the signal from the single-photon counter is sent to a time-resolved single-photon counter (Picoharp300, PicoQuant) and the time-bin is set to be 512 ps.

\subsection*{DC and AC electronics for measurements}
The DC voltages are generated using an arbitrary waveform generator (M3202A, Keysight), and the voltages are amplified using amplifiers (EVAL-ADHV4702-1CPZ, Analog Devices). The microwave voltages of up to 16 GHz ($>$ 25 GHz for a single channel) are generated using another arbitrary waveform generator (M8195A, Keysight), and the microwave voltages are amplified using an amplifier (ZVE-3W-183+, Mini Circuits). Additionally, the microwave power at the sample is estimated using a network analyzer (P9373A, Keysight). The loss inside the cryostat is estimated from the reflection coefficient, $S_{11}$, inside the cryostat. Assuming equal energy loss on the way to the sample and back, half of $S_{11}$ corresponds to the loss between the input port and sample. 
\newline

\section*{Data availability}
Data available in a public repository that issues datasets with DOIs.
	
\section*{Acknowledgements}
This work was supported by a Japan Science and Technology Agency (JST) Moonshot R$\&$D grant (JPMJMS2062) and a JST CREST grant (JPMJCR1773). We also acknowledge the Ministry of Internal Affairs and Communications (MIC) for funding, research and development for construction of a global quantum cryptography network (JPMI00316), and the Japan Society for the Promotion of Science (JSPS) Grants-in-Aid for Scientific Research (20H05661, 20K20441).
\section*{Authorship contribution}
H. Kurokawa and Y. S. designed the experiments. H. Kurokawa wrote the manuscript. K. W. and S. N performed the experiments. T. M and H. Kato fabricated the electrical circuit. H. Kosaka supervised the project.

\bibliographystyle{apsrev4-1-noarXiv}
%

\end{document}


\title{Supplementary Material for: Coherent Electric Field Control of Orbital State of a Neutral Nitrogen-Vacancy Center}
	
	
	\author{Hodaka Kurokawa}
	\email[E-mail: ]{kurokawa-hodaka-hm@ynu.ac.jp}
	\address{Quantum Information Research Center, Institute of Advanced Sciences, Yokohama National University, 79-5 Tokiwadai, Hodogaya, Yokohama 240-8501, Japan}
	\author{Keidai Wakamatsu}
	\address{Department of Physics, Graduate School of Engineering Science, Yokohama National University,  79-5 Tokiwadai, Hodogaya, Yokohama 240-8501, Japan }
	\author{Shintaro Nakazato}
	\address{Department of Physics, Graduate School of Engineering Science, Yokohama National University,  79-5 Tokiwadai, Hodogaya, Yokohama 240-8501, Japan }
	\author{Toshiharu Makino}
	\address{Advanced Power Electronics Research Center, National Institute of Advanced Industrial Science and Technology, 1-1-1 Umezono, Tsukuba, Ibaraki, 305-8568, Japan }
	\author{Hiromitsu Kato}
	\address{Advanced Power Electronics Research Center, National Institute of Advanced Industrial Science and Technology, 1-1-1 Umezono, Tsukuba, Ibaraki, 305-8568, Japan }
	\author{Yuhei Sekiguchi}
	\address{Quantum Information Research Center, Institute of Advanced Sciences, Yokohama National University, 79-5 Tokiwadai, Hodogaya, Yokohama 240-8501, Japan}
	\author{Hideo Kosaka}
	\email[E-mail: ]{kosaka-hideo-yp@ynu.ac.jp}
	\address{Quantum Information Research Center, Institute of Advanced Sciences, Yokohama National University, 79-5 Tokiwadai, Hodogaya, Yokohama 240-8501, Japan}
	\address{Department of Physics, Graduate School of Engineering Science, Yokohama National University,  79-5 Tokiwadai, Hodogaya, Yokohama 240-8501, Japan }

	\maketitle

\beginsupplement
\section{Hamiltonian of NV$^0$}
When the magnetic fields and strain are zero, the Hamiltonian of NV$^0$ can be written in basis \(\ket{\pm}_{\text{o}} = \mp(\ket{e_x}_{\text{o}} \pm i\ket{e_y}_{\text{o}})\) as follows:
\begin{equation}
	H_0^{(\pm)}/h = 2\lambda \hat{L}_z \hat{S}_z,
\end{equation}
where $h$ is the Plank's constant, $\lambda$ is the spin-orbit interaction parameter, $\hat{L}_z=\sigma_z$, $\hat{L}_{\pm}=\ket{\pm}_{\text{o}}\bra{\mp}_{\text{o}}$ are the orbital operator in the $\ket{\pm}_{\text{o}}=\mp1/\sqrt{2}(\ket{e_x}_{\text{o}}\pm i\ket{e_y}_{\text{o}})$ basis, $\ket{e_x}_{\text{o}}$ and $\ket{e_y}_{\text{o}}$ are the strain eigenstates, and $S_z=(1/2)\sigma_z$ is the spin operator for the 1/2 spin.
Because NV$^0$ has the same wave function symmetry as the excited state of NV$^-$ due to the dynamic Jahn-Teller effect (the effective C$_{3v}$ symmetry), the effects of the strain and electric fields on $\ket{e_{x,y}}_{\text{o}}$ are considered to be similar to that in the excited state of NV$^-$ \cite{ Gali2009,Maze2011,Zhang2011}. Therefore, the discussion on the excited state of NV$^-$ in Ref. \cite{Maze2011} is applied on NV$^0$ as outlined below.
In the $\ket{e_{x,y}}_{\text{o}}$ basis, the strain and electric field Hamiltonian can be written as:
\begin{eqnarray}
	H_{\text{strain}}^{(e_{x,y})}/h &=& \epsilon_{A_1} \hat{I} + \epsilon_{E_1} \hat{\sigma}_z - \epsilon_{E_2} \hat{\sigma}_x,\\
	H_{\text{electric}}^{(e_{x,y})}/h &=& d_\parallel E_z \hat{I} + d_\perp E_x \hat{\sigma}_z - d_\perp E_y \hat{\sigma}_x,
\end{eqnarray}
where \(\epsilon_{A_1}\), \(\epsilon_{E_1}\), and \(\epsilon_{E_2}\) represent the strain in the energy dimensions with symmetric indices, \(d_\perp\) denotes the in-plane electric susceptibility, and \(d_\parallel\) represents the electric susceptibility along the NV axis. 
In the $\ket{\pm}_{\text{o}}$ basis: 
\begin{equation}
	\begin{split}
		H_{\text{strain}}^{(\pm)}/h = \epsilon_{A_1} \hat{I} - \epsilon_{E_1} (\hat{L}_+ + \hat{L}_-) \\
		- \epsilon_{E_2} (-i\hat{L}_+ + i\hat{L}_-),\\		
	\end{split}
\end{equation}
\begin{equation}
	\begin{split}
		H_{\text{electric}}^{(\pm)}/h = d_\parallel E_z \hat{I} - d_\perp E_x (\hat{L}_+ + \hat{L}_-) \\
		- d_\perp E_y (-i\hat{L}_+ + i\hat{L}_-).
	\end{split}
\end{equation}
To simplify this analysis, only the subspace of spin in the state \(\ket{\uparrow}_{\text{s}}\) is considered because the strain and electric fields do not directly affect the spin. 

\subsection{Strain and DC electric fields}
First, the state where only the strain is present in the system is considered. For the strain Hamiltonian, the axis of rotation can be redefined for simplicity and expressed as follows:
\begin{equation}
	H_{\text{strain}}^{(\pm)}/h = \epsilon_{A_1} \hat{I} + \epsilon_{\perp} (\hat{L}_+ + \hat{L}_-).
\end{equation}
The electric fields are rewritten as shown below because they may not necessarily align in the same direction as the strain:
\begin{equation}
	H_{\text{electric}}^{(\pm)}/h = d_\parallel E_z \hat{I} + d_\perp E_\perp (\hat{L}_+ + \hat{L}_-) + d_\perp E_{\perp}' (-i\hat{L}_+ + i\hat{L}_-),
\end{equation}
where $E_\perp$ represents the electric fields with the same direction as $\epsilon_{\perp}$, and $E_\perp'$ represents the electric fields perpendicular to $E_z$ and $E_{\perp}'$.
To diagonalize these equations, $H_0+H_\text{strain}+H_\text{electric}$, the energy eigenvalues are obtained as:
\begin{equation}
	E_\pm = (\epsilon_{A_1}+d_{\parallel}E_z)\pm\sqrt{\lambda^2 + (\epsilon_\perp + d_\perp E_\perp)^2 + (d_\perp E_{\perp}')^2}.
\end{equation}

The corresponding eigenvectors without normalization can be written as:
\begin{equation} \label{Supply_eq:eigenstate_strain_e_1}
	\ket{+'}_{\text{o}} = \frac{\lambda + \sqrt{\lambda^2 + (\epsilon_\perp + d_\perp E_\perp)^2 + (d_\perp E_{\perp}')^2}}{\epsilon_\perp + d_\perp E_\perp + i d_\perp E_{\perp}'}\ket{+}_{\text{o}} + \ket{-}_{\text{o}},
\end{equation}

\begin{equation}\label{Supply_eq:eigenstate_strain_e_2}
	\ket{-'}_{\text{o}} = \frac{\lambda - \sqrt{\lambda^2 + (\epsilon_\perp + d_\perp E_\perp)^2 + (d_\perp E_{\perp}')^2}}{\epsilon_\perp + d_\perp E_\perp + i d_\perp E_{\perp}'}\ket{+}_{\text{o}} + \ket{-}_{\text{o}}.
\end{equation}

\subsection{Strain and AC electric fields}

The AC electric fields are applied in such a way that they resonate with the eigenstates of the strain, $\ket{\pm'}_{\text{o}}$ ($\boldsymbol{E}=0$). Therefore, the AC electric fields are treated as a perturbation to the eigenstates of the strain. 

The effects of the electric fields in the basis of \(\ket{\pm'}_{\text{o}}\) with normalized eigenvectors \(\ket{+'}_{\text{o}} = \alpha\ket{+}_{\text{o}} + \beta\ket{-}_{\text{o}}\) and \(\ket{-'}_{\text{o}} = -\beta\ket{+}_{\text{o}} + \alpha\ket{-}_{\text{o}}\) can be expressed as:
\begin{equation}
	\begin{split}
		H_{\text{electric}}^{(\pm')}/h = d_\parallel E_z \hat{I}(t) - d_\perp E_x(t) [(\alpha^2 - \beta^2)(\hat{L}_+ + \hat{L}_-) \\
		+ 2\alpha\beta\hat{L}_z] - d_\perp E_y(t) (-i\hat{L}_+ + i\hat{L}_-),
	\end{split}
\end{equation}
where $\hat{L}_{\pm}$ are the rising (lowering) operators in the $\ket{\pm'}_{\text{o}}$ basis. The equation with the redefined axis after introducing \(E_\perp'' = \sqrt{(\alpha^2 - \beta^2)^2 E_x^2 + E_y^2}\) is :
\begin{equation}
	H_{\text{electric}}^{(\pm')}/h = d_\parallel E_z(t) \hat{I} + d_\perp E_\perp''(t) (\hat{L}_+ + \hat{L}_-) + 2d_\perp E_x(t) \alpha \beta \hat{L}_z
\end{equation}
Subsequently, considering the sinusoidal driving fields, we move to the rotating frame with $H_\text{rot}/h=f_\text{d}\hat{L}_z/2$ from the system Hamiltonian, $\sqrt{\lambda^2+\epsilon_{\perp}^2} \hat{L}_z+H_{\text{electric}}^{(\pm')}/h$. The equation under the rotating wave approximation is
\begin{equation}
	H_{\text{electric}}^{(\pm')}/h = \frac{\Delta}{2} \hat{L}_z + \frac{d_\perp E_\perp''}{2} (\hat{L}_+ + \hat{L}_-).
\end{equation}
Therefore, Rabi oscillation can be observed between the orbital states when $\Delta=2\sqrt{\lambda^2+\epsilon_{\perp}^2}- f_\text{d}\sim0$.

Also, we briefly mention the derivation of the Autler-Townes splitting in the main text. When we set $\Delta=0$, $H_{\text{electric}}^{(\pm')}/h=d_\perp E_\perp'' (\hat{L}_+ + \hat{L}_-)/2$. After the diagonalization of the Hamiltonian, the resultant energy eigenvalues of the dressed states are $\pm d_\perp E_\perp''/2$. The energy splitting between two dressed states is $d_\perp E_\perp''$, corresponding to the Rabi frequency at that driving amplitude. Thus, we can estimate $d_\perp$ from the energy splitting if we know $E_\perp''$.

\section{Estimation of the applied electric field}

\begin{figure}
	\begin{center}
		\includegraphics[width=80 mm]{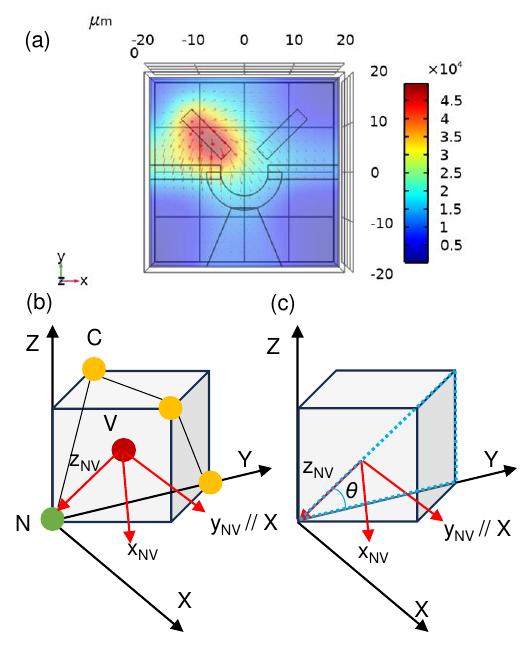}
	\end{center}
	\caption{(a) Electric field distribution around the electrode simulated using FEM software using an applied voltage of 1 V. The unit of the color bar is V/m. (b) Relationship between the laboratory frame and frame of the NV center. C, N, and V correspond to the carbon atom, nitrogen atom, and vacancy, respectively. The direction of the NV center is estimated from the measurement of the optically-detected magnetic resonance. Only two of the four NV axes can be distinguished through measurement; however, the distinction between the two types of axes only affects the sign of the electric field. Therefore, the NV axis is defined as illustrated in the figure. $z_\text{NV}$ is along the NV axis, and $x_\textrm{NV}$ and $y_\textrm{NV}$ are defined following the definition of $\ket{e_{x,y}}$ in Ref. \cite{Maze2011}. (c) $\theta$ is defined as shown in the figure.}
	\label{fig:E_field}
\end{figure}

The electric fields and applied voltage are estimated based on a simulation of the electric field distribution using the finite element method (FEM) (COMSOL Multiphysics, COMSOL). The electric fields at the NV center are estimated to be ($E_X$, $E_Y$, $E_Z$) = (12497.6 V/m, -26122.3 V/m, -7973.57 V/m) per 1 V for the electrode used for DC voltage application, where $E_X$, $E_Y$, $E_Z$ are the electric fields in the laboratory frame, as shown in Fig. \ref{fig:E_field} (a). For the electrode used for ac voltage application, ($E_X$, $E_Y$, $E_Z$) = (13763.6 V/m, -18844.1 V/m, -1079.8 V/m) per 1 V. The electric fields in the NV frame, $E_x$, $E_y$, and $E_z$, are defined as $E_x = E_Y\text{sin}\theta - E_Z\text{cos}\theta$, and $E_y = E_X$, $E_z = -E_Y\text{cos}\theta -E_Z\text{sin}\theta$, where $\text{sin}\theta$ = $\sqrt{1/3}$ and $\text{sin}\theta$ = $\sqrt{2/3}$ (Fig. \ref{fig:E_field} (b)(c)). Therefore, ($E_x$, $E_y$, $E_z$) = (-8571.3 V/m, 12497.6 V/m, 25932.3 V/m) per 1 V for the electrode used for DC voltage application, and ($E_x$, $E_y$, $E_z$) = (-9998.0 V/m, 13763.6 V/m, 16009.6 V/m) per 1 V for the electrode used for ac voltage application. The applied AC voltage is estimated using the relation $V=\sqrt{PR}$, where $V$ is the root-mean-square (RMS) voltage, $P$ is the microwave power, and $R$ is the resistance. When 1 $\mu$W microwave is applied to the electrode at the end of the 50 $\Omega$ transmission line, $V$ = 7.1 mV. Assuming that the end of the electrode is an approximate open circuit (Z = $\infty$), the voltage amplitude at the electrode is $7.1\times\sqrt{2}\times2=20$ mV. The factor $\sqrt{2}$ is introduced to convert the RMS voltage to the voltage amplitude, wherein a factor of 2 is used because the circuit is open at the end of the electrode.

\section{Comparison between electric field control with magnetic field control}

In this section, we compare the efficiency of the electric field orbital control with that of the magnetic field spin  control to support the argument in the main text that the electric field orbital control can be performed with three orders of magnitude less power than that required for the magnetic field spin control.

If we assume that the NV center is 5 $\mu$m away from the electrode, The Rabi frequency for the spin is $\Omega_{\textrm{s}}=g\mu_{\textrm{B}}B_x\hat{S}_x/\hbar$, where $g$ is the electron $g$-factor, $\mu_{\textrm{B}}$ is the Bohr magneton, $B_x$ is $x$-axis magnetic field, and $\hat{S}_x$ is the spin operator. When $\Omega_{\textrm{s}}$ is set to be 100 MHz, $B_x$ must be 34.8 (69.7) G for NV$^-$, spin-1 (NV$^0$, spin-1/2). From Ampere’s law, the required current is 87.5 (175) mA, corresponding to 382 (1530) mW for a 50 $\Omega$ transmission line ($P = I^2Z$), where $P$ is power, $I$ is current, $Z$ is impedance.
On the other hand, the Rabi frequency for orbital state can be written as $\Omega_\textrm{o}=d_{\perp}E_{\perp}$, where $d_{\perp}$ is the electric field susceptibility ($\sim$1000 kHz/V cm$^{-1}$), and $E_{\perp}$ is the electric field. When $\Omega_\textrm{o}$ is 100 MHz, $E_{\perp}$=100 V/cm. When NV is 5 $\mu$m away from the both electrodes (in between the signal and ground), the required voltage is 0.1 V, corresponding to 200 $\mu$W in the 50 $\Omega$ transmission line. Ideally, therefore, power suppression is expected to be more than three orders of magnitude.

In our samples, experimentally, $\sim$500 $\mu$W is required to control the orbital state of NV$^0$ with the Rabi frequency of $\sim$90 MHz, while $\sim$300 mW was needed to control the spin state of NV$^-$ with the Rabi frequency of $\sim$90 MHz. The  required power for the magnetic field control is in good agreement with the above estimation. In contrast, the power required for electric field control is larger than our estimation, which may be due to the displacement and misorientation of NV from the ideal position between the electrodes.

\bibliographystyle{apsrev4-2}
%